\documentclass[12pt]{iopart}
\usepackage{graphicx}

\begin{document}

\title{Localized energy associated with Bianchi-Type VI universe in $f(R)$ theory of gravity}

\author{M. Korunur\footnote[1]{E-mail: muratkorunur@yahoo.com}}

\address{Electric and Energy Department, Tunceli Vocational School, Tunceli University\\
Tunceli-62000, Turkey}

\begin{abstract}
In the present work, focusing on one of the most popular problems in
modern gravitation theories, we consider generalized
Lanndau-Liftshitz energy-momentum relation to calculate energy
distribution of the Bianchi-Type VI spacetime in $f(R)$ gravity.
Additionally, the results are specified by using some well-known $f(R)$-gravity models.
\end{abstract}

\section{Introduction}
Gravitational energy-momentum localization problem is one of the
most popular in gravitation theories and it still remains unsolved.
Einstein is known as the first scientist who worked on
energy-momentum pseudotensors\cite{einstein} and different energy
momentum prescriptions\cite{tlmn,pp,ll,bt,gold,wein,habib1} were put
forwarded after his studies. All energy-momentum formulations except
for the M{\o}ller prescription\cite{m1} were restricted to make
computations in cartesian coordinates. In 1990, Virbhadra and his
collaborators re-opened the energy-momentum localization
problem\cite{v1,v2,v3,v4,v5,v6,v7} and after those pioneering papers
great numbers of work have been
prepared\cite{Vagenas,Vagenas2,Vagenas3,Radinschi1,Radinschi2,Xulu,sharif1,Salt1,Salt2,Salt3,Oktay,sezgin1,sezgin2}
by considering different energy momentum complexes and spacetime
models.

Recently, modified gravitation theories especially $f(R)$ gravity
which extends the general theory of relativity have also been taken
into account by many scientists to discuss gravitational puzzles
again\cite{edd,buc,Carroll,cap1,n1,n2,alle,meng,mul,sta,hol,sh2,sh3,sh4,j1,far,Askin,Habib,Msalti}.
The $f(R)$-gravity should be defined by modifying the
Einstein-Hilbert action:
\begin{equation}
S=-\frac{1}{2\kappa}\int\sqrt{-g}f(R)d^4x+S_m\label{eh1}
\end{equation}
where $\kappa=8\pi G$, $g$ represents the determinant of the metric
tensor, $f(R)$ denotes a general function of Ricci scalar and $S_m$
is the matter part of action\cite{duv}. It is known that the Ricci
curvature scalar is given by
\begin{equation}
R=g^{\mu\nu}R_{\mu\nu}\label{ricci},
\end{equation}
where $R_{\mu\nu}$ is the Ricci tensor related with the Riemann
tensor, i.e. $R_{\mu\nu}=R^{\lambda}_{\mu\lambda\nu}$, given below:
\begin{equation}\label{rieman}
R^{\lambda}_{\mu\nu\sigma}=\partial_\nu\Gamma^{\lambda}_{\mu\sigma}-\partial_\sigma\Gamma^{\lambda}_{\mu\nu}+\Gamma^{\eta}_{\mu\sigma}\Gamma^{\lambda}_{\eta\nu}-\Gamma^{\eta}_{\mu\nu}\Gamma^{\lambda}_{\eta\sigma},
\end{equation}
and $\Gamma^{\lambda}_{\mu\sigma}$ is known as the Christoffel
symbols:
\begin{equation}\label{chris}
\Gamma^{\lambda}_{\mu\sigma}=\frac{1}{2}g^{\lambda\beta}\left(\partial_{\sigma}g_{\mu\beta}+\partial_{\mu}g_{\sigma\beta}-\partial_{\beta}g_{\mu\sigma}\right).
\end{equation}
Now, varying equation (\ref{eh1}) with respect to the metric tensor
yields the following field equation
\begin{equation}
F(R)R_{\mu\nu}-\frac{1}{2}f(R)g_{\mu\nu}-[\nabla_{\mu}\nabla_{\nu}-g_{\mu\nu}\nabla_{\alpha}\nabla^{\alpha}]F(R)=\kappa
T_{\mu\nu}.
\end{equation}
Here, it has been defined that $F(R)\equiv\frac{df(R)}{dR}$ and
$\nabla_{\mu}$ represents the covariant derivative.  After
construction for the vacuum case, i.e. $T=0$, the corresponding
field equation transforms to the following form
\begin{equation}\label{2}
F(R)R-2f(R)+3\nabla_{\alpha}\nabla^{\alpha}F(R)=0.\label{ad1}
\end{equation}
It can be easily seen that for any constant curvature scalar
equation (\ref{ad1}) becomes
\begin{equation}
F(R_0)R_0-2f(R_0)=0,
\end{equation}
here we have used that $R=R_0=constant$. In non-vacuum case, the
constant curvature scalar condition is described by
\begin{equation}
F(R_0)R_0-2f(R_0)=\kappa T.
\end{equation}

Making use of the generalized Landau-Liftshitz prescription for the
Schwarzschild-de Sitter universe, Multam\"{a}ki et al.\cite{mult}
calculated energy distribution for some well known $f(R)$ gravity
models including constant curvature scalar. Later, Amir and
Naheed\cite{j2} considered a spatially homogeneous rotating
spacetime solution of $f(R)$ gravity to calculate Landau-Liftshitz
energy density. Moreover, using some well-known $f(R)$ theory
suggestions, Salti et al.\cite{murat} also discussed energy-momentum
localization problem for G\"{o}del-Type metrics. These studies
motivate us to discuss energy-momentum problem for another
background in $f(R)$-gravity and extend those works.

The paper is organized as follows. In the second section, we give a
brief information about the Landau-Liftshitz distribution in $f(R)$
gravity for the Bianchi-VI type spacetime. Next, in the third
section, we calculate energy density considering Landau-Liftshitz
distribution for some specific $f(R)$ models. Finally, we devote the
last section to discussions.

\section{Generalized Landau-Liftshitz Prescription in Bianchi-Type VI Spacetime}
The generalized Landau-Liftshitz distribution is given by\cite{mult}
\begin{equation}\label{a3}
\tau^{\mu\nu}=F(R_0)\tau_{LL}^{\mu\nu}+\frac{1}{6\kappa}[F(R_0)R_0-f(R_0)]\frac{\partial}{\partial
x^{\gamma}}(g^{\mu\nu}x^{\gamma}-g^{\mu\gamma}x^{\nu}),
\end{equation}
where $\tau_{LL}^{\mu\nu}$ is the Landau-Lifshitz energy-momentum of
general relativity and defined by
\begin{equation}\label{a4}
\tau_{LL}^{\mu\nu}=(-g)(T^{\mu\nu}+t_{LL}^{\mu\nu})
\end{equation}
with
\begin{eqnarray}\label{a5}
t_{LL}^{\mu\nu}&=&\frac{1}{2\kappa}\left[(2\Gamma_{\alpha\beta}^{\gamma}\Gamma_{\gamma\delta}^{\delta}-\Gamma_{\alpha\delta}^{\gamma}\Gamma_{\beta\gamma}^{\delta}
-\Gamma_{\alpha\gamma}^{\gamma}\Gamma_{\beta\delta}^{\delta})(g^{\mu\alpha}g^{\nu\beta}-g^{\mu\nu}g^{\alpha\beta})\right.\nonumber\\
&&+g^{\mu\alpha}g^{\beta\gamma}(\Gamma_{\alpha\delta}^{\nu}\Gamma_{\beta\gamma}^{\delta}+\Gamma_{\beta\gamma}^{\nu}\Gamma_{\alpha\delta}^{\delta}
-\Gamma_{\gamma\delta}^{\nu}\Gamma_{\alpha\beta}^{\delta}-\Gamma_{\alpha\beta}^{\nu}\Gamma_{\gamma\delta}^{\delta})\nonumber\\
&&+g^{\nu\alpha}g^{\beta\gamma}(\Gamma_{\alpha\delta}^{\mu}\Gamma_{\beta\gamma}^{\delta}+\Gamma_{\beta\gamma}^{\mu}\Gamma_{\alpha\delta}^{\delta}
-\Gamma_{\gamma\delta}^{\mu}\Gamma_{\alpha\beta}^{\delta}-\Gamma_{\alpha\beta}^{\mu}\Gamma_{\gamma\delta}^{\delta})\nonumber\\
&&\left.+g^{\alpha\beta}g^{\gamma\delta}(\Gamma_{\alpha\gamma}^{\mu}\Gamma_{\beta\delta}^{\nu}-\Gamma_{\alpha\beta}^{\mu}\Gamma_{\gamma\delta}^{\nu})\right].
\end{eqnarray}
Consider $00$-component of equation (\ref{a3}) gives energy density
associated with the universe and it can be written as given
below\cite{mult}:
\begin{equation}\label{6}
\tau^{00}=F(R_0)\tau_{LL}^{00}+\frac{1}{6\kappa}[F(R_0)R_0-f(R_0)](\frac{\partial}{\partial
x^{i}}g^{00}x^{i}+3g^{00}).
\end{equation}

In the canonical cartesian coordinates, the homogenous Bianchi-Type
VI spacetime is defined by the following
line-element\cite{fagundes}:
\begin{equation}
ds^{2}=dt^{2}-dx^{2}-e^{2(A-1)x}dy^{2}-e^{2(A+1)x}dz^{2}, \label{1}
\end{equation}
where $A$ is a constant with $0\leq A\leq 1$. The metric tensor
$g_{\mu\nu}$, its form $g^{\mu\nu}$ and $\sqrt{-g}$ for the
Bianchi-Type VI model can be written, respectively, as:
\begin{equation}
g_{\mu\nu}=(1,-1,-e^{2(A-1)x},-e^{2(A+1)x}),
\end{equation}
\begin{equation}
g^{\mu\nu}=(1,-1,-e^{2(1-A)x},-e^{-2(A+1)x}), \label{3}
\end{equation}
\begin{equation}
\sqrt{-g}=e^{2Ax}. \label{4}
\end{equation}
Next, the nonvanishing component of Christoffel symbols are
calculated as
\begin{eqnarray}
\Gamma_{22}^{1}&=&(A-1)e^{2(A-1)x},\nonumber\\
\Gamma_{33}^{1}&=&-(A+1)e^{2(A+1)x},\nonumber\\
\Gamma_{12}^{2}&=&\Gamma_{21}^{2}=(A-1),\nonumber\\
\Gamma_{13}^{3}&=&\Gamma_{31}^{3}=(A+1).\label{5}
\end{eqnarray}
Using the above results, the surviving components of Ricci tensor
become
\begin{eqnarray}
R_{11}&=&-2(A^2+1),\nonumber\\
R_{22}&=&2A(1-A)e^{2(A-1)x},\nonumber\\
R_{33}&=&-2A(1+A)e^{2(A+1)x}.
\end{eqnarray}
Additionally, the constant value of Ricci scalar is
\begin{equation}
R=R_0=6A^2+2.
\end{equation}
Making use of above calculations, the non-vanishing components of
$t_{LL}^{\mu\nu}$ are found as
\begin{eqnarray}
t_{LL}^{00}&=&\frac{1}{\kappa}(1-5A^2),\nonumber\\
t_{LL}^{11}&=&\frac{1}{\kappa}(1-A^2),\nonumber\\
t_{LL}^{22}&=&\frac{1}{\kappa}\left[\frac{(1+A)^2}{e^{2(A-1)x}}\right],\nonumber\\
t_{LL}^{22}&=&\frac{1}{\kappa}\left[\frac{(A-1)^2}{e^{2(A+1)x}}\right].
\end{eqnarray}
Also, the non-zero components of $\tau_{LL}^{\mu\nu}$ are calculated
as:
\begin{eqnarray}
\tau_{LL}^{00}&=&-\frac{8A^2e^{4Ax}}{\kappa},\nonumber\\
\tau_{LL}^{22}&=&\frac{2}{\kappa}(A+1)^2e^{2(A+1)x},\nonumber\\
\tau_{LL}^{33}&=&\frac{2}{\kappa}(A-1)^2e^{2(A-1)x}.
\end{eqnarray}
Consequently, in the f(R)-gravity, one can easily write down the
generalized form of Landau-Liftshitz energy distribution as given
below
\begin{equation}
\tau^{00}=-\frac{1}{2\kappa}\left\{\left[R_0F(R_0)-f(R_0)\right]-16A^2e^{4Ax}F(R_0)\right\}\label{t0},
\end{equation}
and we also have
\begin{eqnarray}
\tau^{0i}&=&\frac{1}{6\kappa}\left[f(R_0)-R_0F(R_0)\right],\qquad(i=1,2,3),\nonumber\\
\tau^{11}&=&\frac{2}{3\kappa}\left[f(R_0)-R_0F(R_0)\right],\nonumber\\
\tau^{22}&=&\frac{e^{2(1-A)x}}{3\kappa}\left\{\left[2+(1-A)x\right]f(R_0)\right.\nonumber\\&&\left.+\left[6(A+1)^2e^{4Ax}+(Ax-x-2)R_0\right]F(R_0)\right\},\nonumber\\
\tau^{33}&=&\frac{e^{-2(1+A)x}}{3\kappa}\left\{\left[2-(1+A)x\right]f(R_0)\right.\nonumber\\&&\left.+\left[6(A-1)^2e^{4Ax}+(Ax+x-2)R_0\right]F(R_0)\right\}.
\end{eqnarray}

\section{Energy in specific $f(R)$ Models}
There are many suggested models in the $f(R)$ theory of
gravity\cite{capoziello}. In this section of the study, we mainly
consider five different well-known models to calculate energy
momentum distribution associated with Bianchi-Tpe VI spacetime
exactly.

\begin{itemize}
    \item The first model\cite{staro,noakes} is described in a polynomial
form:
\end{itemize}
\begin{equation}
f_{1st}(R)=R+\xi R^2,
\end{equation}
where $\xi$ denotes a positive real number.

\begin{itemize}
    \item The second model\cite{Faulkner} is given by
\end{itemize}
\begin{equation}
f_{2nd}(R)=R-\frac{\epsilon^4}{R},
\end{equation}
where $\epsilon$ is a constant parameter. This model is known also
as the dark energy model of $f(R)$-gravity.

\begin{itemize}
    \item The next model\cite{Nojiri} is defined as
\end{itemize}
\begin{equation}
f_{3th}(R)=R-pR^{-1}-qR^2,
\end{equation}
with $p$, $q$ are constant.

\begin{itemize}
    \item Another one is given by the following definition\cite{Nojiri2}:
\end{itemize}
\begin{equation}
f_{4th}(R)=R-p\ln\left(\frac{|R|}{\sigma}\right)+(-1)^{n-1}qR^n.
\end{equation}
Here $n$ represents an integer and $p $, $q$, $\sigma$ are constant
parameters.

\begin{itemize}
    \item The final model is known as the chameleon model and it is given
by\cite{Faulkner}
\end{itemize}
\begin{equation}
f_{5th}(R)=R-(1-m)\lambda^2
\left(\frac{R}{\lambda^2}\right)^m-2\Lambda,
\end{equation}
where $\Lambda$ denotes the cosmological constant, $m$ shows an
integer and $\lambda$ is a constant parameter.

For suitable choices of above constants, all of the $f(R)$ models
mentioned above can define the general relativity exactly. Now,
considering the above $f(R)$ gravity models and equation (\ref{t0}),
one can obtain the following energy densities:
\begin{equation}
\tau_{1st}^{00}=\frac{1}{\kappa}\left\{2(3A^2+1)^2\xi-8A^2e^{4Ax}\left[4\xi(3A^2+1)+1\right]\right\},
\end{equation}
\begin{equation}
\tau_{2nd}^{00}=\frac{1}{2\kappa(3A^2+1)^2}\left\{\epsilon^4+A^2\left[3\epsilon^4-16e^{4Ax}(3A^2+1)^2+4\epsilon^4e^{4Ax}\right]\right\},
\end{equation}
\begin{eqnarray}
\tau_{3th}^{00}&=&\frac{1}{2\kappa(3A^2+1)^2}\left\{p(1+3A^2-4A^2e^{4Ax})\right.\nonumber\\
&&\left.+16Ae^{4Ax}(3A^2+1)^2\left[4q(3A^2+1)-1\right]-4q(3A^2+1)^4\right\},
\end{eqnarray}
\begin{eqnarray}
\tau_{4th}^{00}&=&\frac{1}{2\kappa}\left\{(-1)^nq(1-n)(6A^2+2)^n-p\right.\nonumber\\
&&\left.-16A^2e^{4Ax}\left[1-\frac{p+2nq(-1)^n(3A^2+1)^n}{2(3A^2+1)-p\ln\left(\frac{6A^2+2}{\sigma}\right)}\right]\right\},
\end{eqnarray}
\begin{eqnarray}
\tau_{5th}^{00}&=&\frac{1}{2\kappa}\left\{2\Lambda-16A^2e^{4Ax}\right.\nonumber\\
&&\left.-\left(\frac{6A^2+2}{\lambda^2}\right)^m\frac{\lambda^2(m-1)[1-m+A^2(3-3m+8me^{4Ax})]}{3A^2+1}\right\}.
\end{eqnarray}

\section{Concluding Remarks}
Considering Bianchi-Type VI spacetime representation and some
popular models of $f(R)$ gravity with constant Ricci curvature
scalar, we have mainly evaluated the Landau-Liftshitz energy
distribution. All of the calculations have been performed in
cartesian coordinates. We have found that the energy distribution of
Bianchi-Type VI model in $f(R)$ gravity as below:
\begin{equation}
\tau^{00}=-\frac{1}{2\kappa}\left\{\left[R_0F(R_0)-f(R_0)\right]-16A^2e^{4Ax}F(R_0)\right\}.
\end{equation}
Assuming $A_{min}=0$, one can see that the energy momentum
distributions transform into the following forms:
\begin{equation}
\tau_{1st(A=0)}^{00}=\frac{2\xi}{\kappa},
\end{equation}
\begin{equation}
\tau_{2nd(A=0)}^{00}=\frac{\epsilon^4}{2\kappa},
\end{equation}
\begin{equation}
\tau_{3th(A=0)}^{00}=\frac{p-4q}{2\kappa},
\end{equation}
\begin{equation}
\tau_{4th(A=0)}^{00}=\frac{1}{2\kappa}\left[(-2)^nq(1-n)-p+p\ln\left(\frac{2}{\sigma}\right)\right],
\end{equation}
\begin{equation}
\tau_{5th(A=0)}^{00}=\frac{1}{2\kappa}\left[2\Lambda+2^m(1-m)^2\lambda^2(1-m)\right].
\end{equation}
It is seen that all energy distributions are constant. Therefore, it
can be generalized that
\begin{equation}
\tau_{All(A=0)}^{00}=constant
\end{equation}
On the other hand, in case of $A=A_{max}=1$, the energy momentum
distributions for all models do not have constant values as
expected. Moreover, when we take $f(R)=R_0$ in equation (\ref{t0})
it can be concluded that
\begin{equation}
\tau_{GR}^{00}=-\frac{8}{\kappa}A^2e^{4Ax}.
\end{equation}

\end{document}